\documentclass[twocolumn, prl, superscriptaddress,notitlepage]{revtex4-2}
\usepackage{graphicx}% Include figure files
\usepackage{dcolumn}% Align Table columns on decimal point
\usepackage{bm}% bold math
\usepackage[usenames,dvipsnames]{color}
\usepackage[most]{tcolorbox}
\usepackage{multirow}
\usepackage{gensymb}
\usepackage[normalem]{ulem}
\usepackage{CJK}
\usepackage{comment}
\usepackage[colorlinks, linkcolor=blue,anchorcolor=blue,citecolor=blue,urlcolor=blue]{hyperref}
\usepackage{amssymb}
\usepackage{pifont}
\usepackage{physics}
\usepackage{natbib}

\begin{document}

%\preprint{APS/123-QED}

\title{First-principles Calculation of the Temperature-dependent Transition Energies in Spin Defects} 
\author{Hao Tang}
\thanks{These authors contributed equally.}
\affiliation{Department of Materials Science and Engineering, Massachusetts Institute of Technology, MA 02139, USA}

\author{Ariel Rebekah Barr}%
\thanks{These authors contributed equally.}
\affiliation{Department of Materials Science and Engineering, Massachusetts Institute of Technology, MA 02139, USA}

\author{Guoqing Wang}%
\thanks{These authors contributed equally.}
\affiliation{
   Research Laboratory of Electronics, Massachusetts Institute of Technology, Cambridge, MA 02139, USA}
\affiliation{
   Department of Nuclear Science and Engineering, Massachusetts Institute of Technology, Cambridge, MA 02139, USA}

\author{Paola Cappellaro}\email{pcappell@mit.edu}
\affiliation{
   Research Laboratory of Electronics, Massachusetts Institute of Technology, Cambridge, MA 02139, USA}
\affiliation{
   Department of Nuclear Science and Engineering, Massachusetts Institute of Technology, Cambridge, MA 02139, USA}
\affiliation{Department of Physics, Massachusetts Institute of Technology, Cambridge, MA 02139, USA}

\author{Ju Li}\email{liju@mit.edu}
 \affiliation{Department of Materials Science and Engineering, Massachusetts Institute of Technology, MA 02139, USA}
\affiliation{
   Department of Nuclear Science and Engineering, Massachusetts Institute of Technology, Cambridge, MA 02139, USA}

\date{\today}% It is always \today, today,
 \begin{abstract}
% I would make this broader, e.g.
Spin qubits associated with color centers are promising platforms for various quantum technologies. However, to be deployed in robust quantum devices, the variations of their intrinsic properties with the external conditions, and in particular temperature, should be known with high precision. 
Unfortunately, a predictive theory on the temperature dependence of the resonance frequency of electron and nuclear spin defects in solids remains lacking. 
In this work, we develop a first-principles method for the temperature dependence of zero-field splitting, hyperfine interaction, and nuclear quadrupole interaction of color centers. As a testbed, we compare our ab-initio calculation results  with experiments in the Nitrogen-Vacancy (NV) center finding good agreement. Interestingly,  we identify the major origin of the temperature dependence as a second-order effect of phonon vibration. 
The method is generally applicable to different color centers and provides a theoretical tool for designing high-precision quantum sensors.  
\end{abstract}

\maketitle

Color centers, the fluorescent lattice defects in insulators, have been intensively studied as solid-state qubits for quantum computing~\cite{pezzagna2021quantum, de2021materials}, quantum communication~\cite{pfaff2014unconditional,childress2006fault}, and quantum sensing~\cite{degen2017quantum,schirhagl2014nitrogen}. 
The temperature (and strain) dependence of the spin defects properties is a critical factor in their performance: their spatio-temporal  fluctuations in the crystal host would result in degraded coherence times, while the sensitivity of the defects to small variations can be exploited in quantum sensing~\cite{degen2017quantum}. A predictive theoretical model accompanied by a robust computational protocol would be invaluable in both mitigating deleterious effects and selecting the best host/defects combinations for quantum sensing. Here, we developed a first-principles method for predicting the temperature dependence of optical, electronic and nuclear spin transition frequencies. 
We benchmark our calculations using the properties of the nitrogen-vacancy (NV) center in diamond. Our first-principle calculations achieve excellent agreement with the temperature dependence of the NV  zero phonon line (ZPL), zero-field splitting (ZFS), hyperfine interaction and nuclear quadrupole interaction.  Crucially, we find that the dominant part of the temperature dependence is from a second-order dynamical phonon effect. 
Our method paves the way for computation-assisted design of novel quantum sensors using  color centers in solids. 

We select the NV center as our testbed since it has been accurately characterized in experiments for its applications to quantum sensing of magnetic fields~\cite{hong2013nanoscale,barry2020sensitivity}, electric fields~\cite{dolde2011electric}, temperature~\cite{wang2018magnetic,fujiwara2021diamond}, pressure\cite{doherty2014electronic,ho2020probing}, and rotation~\cite{jarmola2021demonstration, soshenko2021nuclear}. %, thanks to the sensitive response of their electronic structure to external perturbations.   
While the  temperature dependence of the NV electronic spin resonance frequency~\cite{doherty2014temperature} has been exploited to probe the local temperature with ultra-high spatial resolution in nanothermometry devices~\cite{neumann2013high, kraus2014magnetic, alkahtani2018tin, nguyen2018all, toyli2013fluorescence, neumann2013high, kucsko2013nanometre, plakhotnik2014all, gottscholl2021sub}, it is also detrimental to some quantum devices since its fluctuations can lead to decoherence~\cite{fang2013high}. 

Despite  extensive experimental study on the temperature dependence of both electronic  and nuclear spin  frequencies~\cite{acosta2010temperature,soshenko2020temperature,jarmola2020robust,barson2019temperature,chen2011temperature,acosta2011broadband}, a predictive theoretical method is still lacking. The temperature dependence has been previously attributed to thermal expansion, but  the calculated temperature shifts are far smaller in absolute magnitude  than the experimental values~\cite{ivady2014pressure,acosta2010temperature,chen2011temperature,acosta2011broadband} (in all cases by an order of magnitude). This discrepancy indicates that other effects dominate the temperature dependence. Various explanations have being proposed for such a discrepancy. Doherty $et$ $al.$ proposed that  the dynamical phonon effect might play an important role in the temperature dependence of the ZFS~\cite{doherty2014temperature}. However, in the absence of a method to evaluate the dynamical phonon effect from  first principles,  its contribution to the temperature dependence could not be unambiguously determined. 
To overcome these challenges, here we develop a theoretical model of  the defect energy transitions $\nu$ dependence on the temperature-induced atomic displacement up to second order. We combine a full  calculation of the phonon spectrum including  density functional perturbation theory (DFPT)~\cite{dfpt} with density functional theory (DFT) calculation of the spin-transition energies with the supercell method~\cite{Nieminen2007}.  
%todo: here I wanted to describe the overall calculation workflow before going into details. Maybe you can do a better job.

Given an electro-nuclear spin system, such as the NV center, we aim at calculating the temperature-induced shift of a transition frequency $\nu$ between any two levels that can be probed experimentally (e.g., by Rabi experiments.) The shift in $\nu$ arises 
 from the temperature-induced atomic displacement, as the thermal excitation of electrons is negligible due to the large energy gap in a broad temperature range. 
For a general transition, the atomic displacement effect is described by the energy surfaces of the two levels as a function of atomic coordinates, as shown in Fig.~\ref{fig:illustration}(a). Expressing the atomic configuration by the normal coordinates of phonon modes $\{q_i\}$, the transition frequency is then a function of $(\{q_i\})$: $\nu=[E_2(\{q_i\})-E_1(\{q_i\})]/\hbar$ ($E_1$ and $E_2$ are the energy level 1 and energy level 2, respectively). At finite temperature, $\{q_i\}$ have both thermal and quantum fluctuations. As all electron and nuclear spin transition frequencies are at least three orders of magnitude smaller than the typical phonon frequency~\cite{doherty2012theory}, the measured $\nu$ is a statistical average of the phonon mode distribution:
\begin{equation}
    \langle\nu\rangle = \nu_0+\sum_i\left[(\frac{\partial\nu}{\partial q_i})_{0}\langle q_i\rangle+\frac{1}{2}(\frac{\partial^2\nu}{\partial q_i^2})_{0}\langle q_i^2\rangle \right]+O(q^3).
    \label{TaylorExpansionAverage}
\end{equation}
The first-order term represents the change of atomic equilibrium positions, corresponding to  thermal expansion. Here we emphasize that the first-order term is typically smaller than the second-order term as $\langle q_i\rangle$ originates from the weak phonon anharmonicity. Then, what appears to be a ``first-order term'' in Eq.~(\ref{TaylorExpansionAverage}) is actually the product of first-order and third-order terms, as a purely harmonic system will have zero thermal expansion with $\langle q_i\rangle=0$. We evaluate the first-order term through the quasiharmonic approximation~\cite{togo2010first} as a static lattice expansion effect. The temperature-dependent lattice parameter $a(T)$  of nitrogen-rich diamond is reproduced from the previous experiments~\cite{diamond_thermal}, and $\nu (a)$ is calculated for different $a$ at equilibrium atomic configuration. The thermal expansion contribution to the spectral drift is then:
\begin{equation}
    (\frac{\partial\nu}{\partial T})_{\rm quasiharmonic} = \frac{\partial \nu}{\partial a} a(T)\alpha (T),
    \label{ThermalExpansion}
\end{equation}
where  $\alpha(T)$ is the linear thermal expansion coefficient at temperature $T$.
 
The dominant second-order term represents  the atomic vibration around the equilibrium positions  caused by phonon excitation at finite temperature. 
The second-order term is evaluated by the dependence of the spin-transition energy $\nu$ on the phonon occupation number $n_i = 0, 1, 2, \cdots$ ($i$ is the vibrational mode index), 
which  affects $\langle q_i^2\rangle$ in 
Eq.~\eqref{TaylorExpansionAverage}.  By evaluating $\langle q_i^2\rangle$ for a quantum harmonic oscillator (harmonic phonon theory) at thermal equilibrium, 
%Eq.~\eqref{eq:td} can also be derived directly from Eq.~\eqref{TaylorExpansionAverage}, since 
$\langle q_i^2\rangle = \frac{\hbar}{M_i\omega_i}(\langle n_i\rangle+\frac{1}{2}) = \frac{\hbar}{M_i\omega_i}\left(\frac{1}{e^{\hbar\omega_i/kT}-1}+\frac{1}{2}\right)$, 
we obtain 
\begin{equation}
    \langle\nu\rangle = \nu_0(a) +\sum_i \frac{1}{2}\frac{\partial^2\nu}{\partial q_i^2}\frac{\hbar}{M_i\omega_i}\left(\frac{1}{e^{\hbar\omega_i/kT}-1}+\frac{1}{2}\right),
    \label{eq:td},
\end{equation}
where $M_i$ is the mode-specific effective mass (conjugate to the $q_i$ definition) in real-space harmonic lattice dynamics expansion.
%$\hbar(2\pi\nu) \approx \frac{ (\Delta k_i) q_i^2}{2} = \frac{M_i \Delta(\omega_i^2) q_i^2}{2}$.
%todo: this needs to be fixed
% more details are moved to SM.
Here we already included the  ``first-order'', thermal expansion term in Eqs.~\eqref{TaylorExpansionAverage}-\eqref{ThermalExpansion}  by taking $\nu_0$ 
not at the zero-temperature lattice constant $a_0$, but at finite-temperature lattice parameter $\nu_0 = \nu_0(a(T))$. 
%We notice that the dynamical phonon effect contributes a finite zero-point vibration term to the spin transition frequency even at zero temperature.
\begin{figure}[t]
\centering
\includegraphics[width=\linewidth]{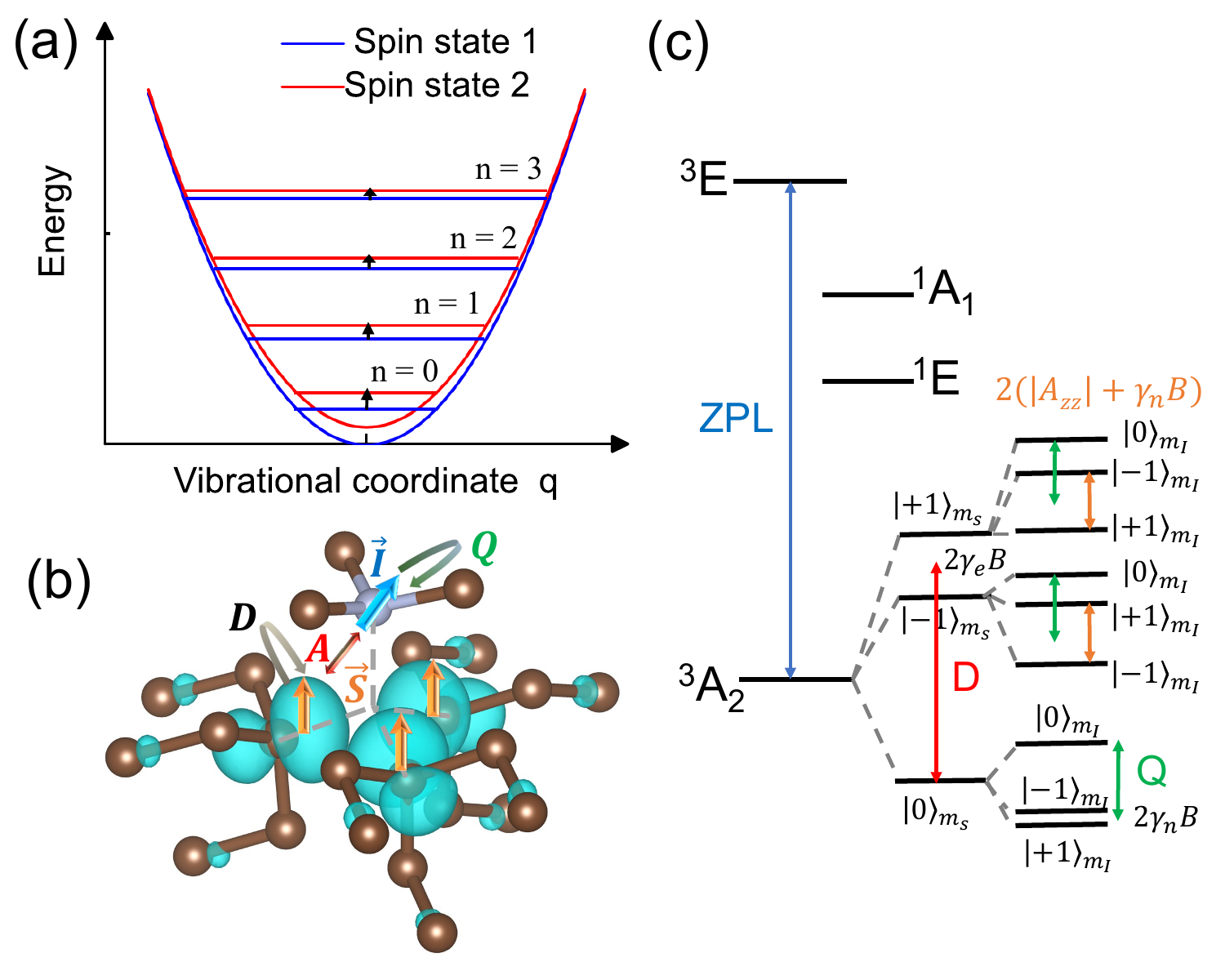}
\caption{(a) Illustration of the second-order phonon effect to spin resonance. (b) Electron and nuclear spin interaction in an NV center. The blue isosurface of electron spin density is from the DFT, and the brown and silver spheres represents carbon and nitrogen atoms, respectively. (c) The energy levels of NV centers.}
\label{fig:illustration}
\end{figure}

We can gain a physical intuition of this result by using the Franck–Condon principle~\cite{lax1952franck} and graphically analyzing the expected frequency shift from Fig.~\ref{fig:illustration}(a). We consider two spin energy levels within vibrational energy levels ($n_i = 0, 1, 2,\cdots$) associated with orthogonal vibrational wavefunctions of the ions. The upper energy surface (red, associated with the higher spin energy) has a slightly softer phonon frequency (on average, at least, for the NV center). Therefore, the spin-transition energy $\nu$ decreases as the phonon number $n_i$ increases. Although the typical radio-frequency and microwave (MHz to GHz) spin control cannot induce direct transitions between different phonon levels $n_i$ ($\sim$THz), these vibration energy levels still affect the average spin transition frequencies by inducing a relatively small shifts in total energy $(n_i+\frac{1}{2})\hbar\Delta\omega_i$, where $\Delta\omega_i = \omega_i^{\rm red}-\omega_i^{\rm blue}$ is the difference in modal frequency between the upper energy surface (red) and the lower energy surface (blue).
%todo: I still do not know what you mean by "integrate spatially" %A revised description is moved to the next paragraph. The spatial integration equation is listed below, which is the overlapping integral of the initial and final state phonon wave function.

%Then, the dynamical phonons decrease the averaged value of the spin-transition energy as:
More specifically, the mean value of the spin transition frequency can be calculated as
\begin{align}
    \langle\nu\rangle &= \sum_{\{n_i\}}\frac{e^{-\sum_i\hbar\omega_i(n_i+\frac{1}{2})/(kT)}}{Z}\left[\nu_0(a)+\sum_i\frac{\Delta\omega_i}{2\pi}(n_i+\frac{1}{2})\right]\nonumber \\
    &= \nu_0(a) + \sum_i \frac{\Delta \omega_i}{2\pi}\left(\frac{1}{e^{\hbar\omega_i/(kT)}-1}+\frac{1}{2}\right)
    \label{eq:stat}.	
\end{align}
As $\Delta \omega_i \ll\omega_i\equiv(\omega_i^{\rm red}+\omega_i^{\rm blue})/2$, the expression can be evaluated by the second order derivative of the spin-transition energy (vertical distance between red and blue) to the normal-mode coordinates $\Delta\omega_i = \frac{\hbar}{2M_i\omega_i}\frac{\partial^2(2\pi\nu)}{\partial q_i^2}$, 
where $(2\pi\hbar\nu)$ is the difference in red and blue potential energy surfaces. %(PES). 
Then the temperature-dependent frequency due to dynamic-harmonic-phonon effect is given by  Eq.~\eqref{eq:td}, which is the form we will adopt for the numerical computations in this paper.  
%The transition rate can also be computed through the Franck–Condon principle~\cite{lax1952franck}, 
%following the horizontal blue line/upward arrow/horizontal red line combinations
%which is proportional to the phonon wavefunction overlaps evaluated by spatial integral $\int\psi_{s_1}^{n*}(q)\psi_{s_2}^{n}(q) dq$ ($\psi_s^n(q)$ is the ionic wave function with spin state $s$ in the ${q_i}$ space).

%We provide proof that this expression corresponds to the central position of the microwave resonance peak measured in experiments in Appendix C.  
Based on the theoretical analysis, we develop the first-principles method to evaluate the quantities in Eq.~\eqref{eq:td}. At first, a full phonon calculation is implemented through the phonopy package~\cite{phonopy} combined with DFPT implemented in Vienna ab-initio simulation package (VASP)~\cite{kresse1996efficient,kresse1999ultrasoft} to derive the frequency $\omega_i$ and effective mass $M_i$ of the full phonon spectrum. The second-order derivatives are then calculated by the finite-differential method:
\begin{equation}
    \frac{\partial^2\nu}{\partial q_i^2} = \frac{1}{\delta q_i^2}[\nu (\delta q_i)+ \nu (-\delta q_i) - 2\nu (0)] + {\cal O}(\delta q_i^2),
    \label{eq:diff}
\end{equation}
where $\nu (q_i)$  is calculated by exerting a small displacement of $\delta q_i$ relative to the relaxed atomic configuration in the DFT calculations (see Supplemental Materials for details~\cite{SOM}).

We simulate the NV center in diamond by a $4\!\times\!4\!\times\!4$ rhombohedral supercell with the single defect at the center. First, we calculate the zero-temperature electronic structures of the NV center for the fully relaxed atomic configuration (Fig.~\ref{fig:illustration}b). The electronic structure calculation employs the DFT and projector-augmented-wave (PAW) method implemented by VASP. The ground state $^3A_2$ contains electron spin $S = 1$ and $^{14}$N nuclear spin $I = 1$, leading to the fine and hyperfine structure shown in Fig.~\ref{fig:illustration}(c). 
The spin state splittings originate from three types of interaction: the electron-electron magnetic dipolar interaction $\boldsymbol{D}$, the hyperfine interaction $\boldsymbol{A}$, and the nuclear quadrupole interaction $\boldsymbol{Q}$~\cite{doherty2013nitrogen}:
\begin{equation}
    \hat{H} = \vec{S}\cdot \boldsymbol{D}\cdot \vec{S}+\vec{S}\cdot \boldsymbol{A}\cdot \vec{I}+\vec{I}\cdot \boldsymbol{Q}\cdot \vec{I}.
    \label{eq:H}
\end{equation}
$\boldsymbol{D}$, $\boldsymbol{A}$, and $\boldsymbol{Q}$ are calculated by the 1$^{st}$-order perturbation theory to the DFT ground state~\cite{ivady2014pressure,gali2008ab,vahtras2002ab,petrilli1998electric,defo2021calculating} (see SM~\cite{SOM} for details). Selecting the $C_3$ axis of the NV center as the $z$-direction, and taking into account that $H_D \gg H_A, H_Q$, the effective Hamiltonian reduces to  ~\cite{doherty2012theory}
\begin{equation}
    \hat{H} = DS_z^2+QI_z^2+A_{zz}S_zI_z,
    \label{eq:h_red}
\end{equation}
where $D= \frac{3}{2}D_{zz}$ and $Q = \frac{3}{2}Q_{zz}$. The temperature variations of $D$, $Q$, and $A_{zz}$ are then calculated by relating them to transitions $\nu$ between  two selected spin states (illustrated in Fig.~\ref{fig:illustration}c).  Similarly, the ZPL's temperature dependence is calculated by Eq.~\eqref{eq:stat} using the $\Delta$SCF method~\cite{PhysRevB.78.075441,gali2009theory,gali2008ab} (see SM~\cite{SOM} for details).

\begin{figure}[b]
\centering
\includegraphics[width=\linewidth]{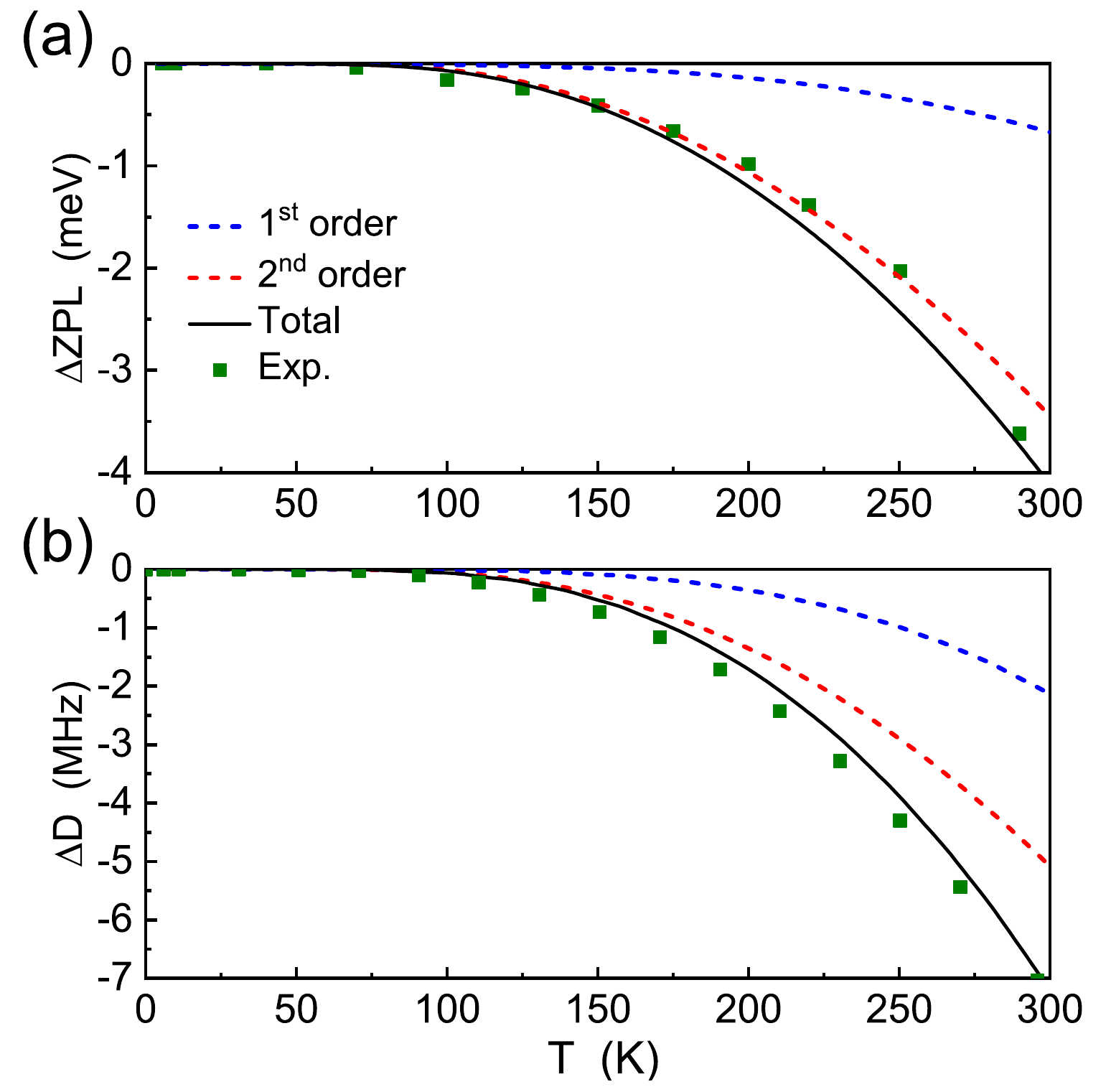} 
\caption{(a) The ZPL and (b) ZFS from simulation and experiment. The blue and red dashed lines represent the first-order (thermal expansion) and second-order (dynamical phonon) effects, respectively. The black line is the total shift from simulation, and the green dots are experimental data points reproduced from Refs.~\cite{doherty2014temperature,chen2011temperature}}
\label{fig:D}
\end{figure}

\begin{figure*}
\centering
\includegraphics[width=1.00\linewidth]{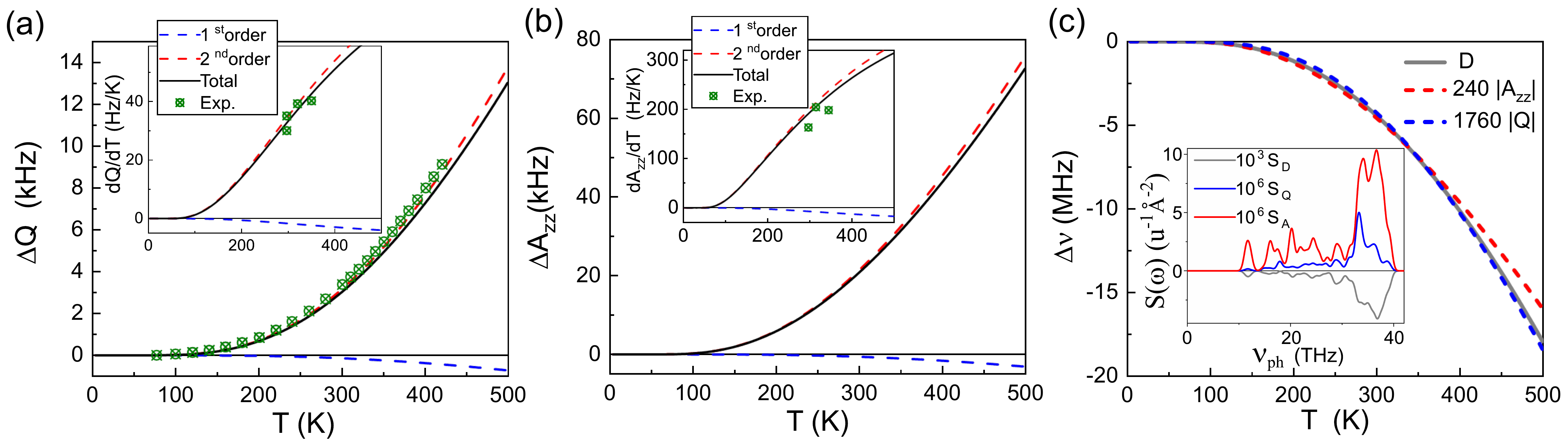}
\caption{(a) Nuclear quadrupole interaction $Q$ and (b) Hyperfine interaction $A_{zz}$ from simulation. The blue, red, and black curves represent the calculated temperature-induced change from thermal expansion effect, 2$^{\rm nd}$-order vibration effect, and their summation, respectively. The insets are the derivative of $Q$ and $A_{zz}$ results, where the green dots are experimental data reproduced from ref.~\cite{doherty2014temperature,soshenko2020temperature,jarmola2020robust,wang2022coherence} (c) Overlap behavior of $D(T)$, $Q(T)$, and $A_{zz}(T)$ from simulation. The inset shows the spectral density $S_D$, $S_Q$, and $S_A$ as defined in Eq.~\eqref{eq:s}, where Gaussian broadening is applied to the $\delta (\omega - \omega_i)$ for each phonon mode with a width of 2 THz.}
\label{fig:AQ}
\end{figure*}

Applying our method to electronic transition, we calculate the temperature dependence of the ZPL and ZFS (Fig.~\ref{fig:D}). The ``first-order'' thermal expansion effect is, as found in previous computational work~\cite{ivady2014pressure,acosta2010temperature}, far smaller than the experimental results for the measured temperature drifts. This is not surprising since the quasi-harmonic ``first-order'' is physically a ``cubic-order'' Gr\"uneisen parameter $\times$ ``first-order'' effect.
Including the second-order dynamical phonon effects corrects the discrepancy. The overall simulation results (summation of the first and second-order) for both quantities are now consistent with prior experimental results. The calculated temperature derivative $\frac{dD}{dT}$ at room temperature is -75.99~kHz/K,  consistent with the widely used experimental value of -74.27~kHz/K~\cite{acosta2010temperature}. Different from Ref.~\cite{doherty2014temperature} where the dynamical phonon effect is fitted by experiment, we evaluate the effect in a parameter-free manner from  first-principles calculations. Our result strongly indicates that the main mechanisms of the temperature dependence of the ZPL and ZFS is the dynamical phonon effect, with a smaller contribution from  the thermal expansion. The temperature shift of the ZFS in NV centers provides promising prospects for nanothermometry, where the magnitude of $\frac{dD}{dT}$ is crucial for its temperature sensitivity~\cite{schirhagl2014nitrogen}. Our method provides a predictive tool to search for different color centers for optimal temperature sensitivity.

We then calculate the temperature dependence of the nuclear spin interaction. The temperature dependence of $^{14}$N-related nuclear spin transition is shown in Figs.~\ref{fig:AQ}(a,b). Both $A_{zz}$ and $Q$ are negative quantities, and their absolute values decrease with  increasing temperature, yielding positive $d Q/dT$ and $d A_{zz}/dT$. The thermal expansion contribution to the temperature shifts has an opposite slope with respect to the experiments. The dynamical phonon effect is more than one order of magnitude larger than the thermal expansion effects, so it corrects both the trend and the magnitude, obtaining $\frac{dQ}{dT}$ and $\frac{dA_{zz}}{dT}$ in good agreement with the previous experiments~\cite{doherty2014temperature,soshenko2020temperature,jarmola2020robust,wang2022coherence}.

$A_{zz}$ is much more sensitive to $T$ than $Q$. The sensitivity of $A_{zz}$ mainly comes from the Fermi contact term that is highly sensitive to the atomic displacement in the dynamical phonon effects according to our calculations. In comparison, the electric field gradient (EFG) is relatively insensitive to atomic displacement, so $Q$ has a smaller temperature shift. Therefore, we expect that the higher sensitivity of $A_{zz}$ than $Q$ is also a general behavior in many other color centers. It has been recently proposed that with $\frac{dA_{zz}}{dQ}>1$, the coherence time of nuclear spin qubits can be robustly protected by at least one order of magnitude through noise decoupling techniques~\cite{wang2022coherence}. The generality of the higher sensitivity of $A_{zz}$ indicates a broad applicability of such a method to various solid state spin defects. 
The relative shift of the hyperfine transition, $\frac{dA_{zz}}{dT}\frac1{A_{zz}}$ approaches $8.9\times 10^{-5} K^{-1}$ at room temperature. While 
%The total shift of transition frequency $\nu = 2A_{zz}$ approaches 40/150 kHz at 300/500 K, which is about 0.8\%/3\% of $\nu$ itself. 
%We notice that although 
the absolute value of ZFS shift  is much larger, the relative frequency change $\frac{dD}{dT}\frac1{D}\approx 2.58\times 10^{-5} $ is smaller than for the hyperfine.
%$\frac{\Delta D}{D}$ is only 0.2\% and 0.5\% at 300 and 500 K, respectively, evidently smaller than the relative change of $A_{zz}$. 
Thus, to operate a quantum sensor based on the nuclear spin it is imperative  to re-calibrate the temperature effects~\cite{fang2013high}.

Previous experimental results had observed a strong, but so-far unexplained, correlation   between $D(T)$ and $Q(T)$~\cite{jarmola2020robust}.
Our first-principles results not only confirm such a correlation and show its validity over a broad temperature range up to 500 K (Fig.~\ref{fig:AQ}c), but also provide approaches to revealing the underlying mechanism of such a correlation. As the dynamical phonon effects are dominant for all these interactions, we investigate the overlap behavior by defining the spectral density of the second-order derivative of $\nu$ as
\begin{equation}
    S_i(\omega ) = \sum_j \frac{1}{M_j}\frac{\partial^2\nu_i}{\partial q_j^2}\delta (\omega -\omega_i),
    \label{eq:s}
\end{equation}
as shown in the inset of Fig.~\ref{fig:AQ}(c). The large intensities of $S_D$, $S_Q$, and $S_A$ at around 32 $\sim$ 38 THz is attributed to the high phonon density of states (DOS)~\cite{zhang2011vibrational} and large second-order derivatives in Eq.~\eqref{eq:s} at that frequency range. The ratio between two temperature-induced frequency shifts is then a weighted average between their spectral density:
\begin{equation}
    \frac{\Delta \nu_1}{\Delta \nu_2} = \frac{\int S_1(\omega )f_{\rm FD}(T, \omega)/\omega d\omega}{\int S_2(\omega )f_{\rm FD}(T, \omega)/\omega d\omega},
\end{equation}
where $f_{\rm FD}$ is the Fermi-Dirac distribution. As the spectral densities $S_Q$ and $S_D$ have a similar shape, the ratio $\Delta Q/\Delta D$ is relatively insensitive to temperature. In comparison, $S_A$ shows a more striking difference, undermining $A_{zz}$'s overlap with $Q$ and $D$. Thus, although $A_{zz}(T)$ also shows an approximate correlation with  $Q$ and $D$, it has a relatively larger deviation with a larger slope in the low-temperature region and a smaller slope in the high-temperature region as shown in Fig.~\ref{fig:AQ}(c). 
We note that the spectral correlation between different interactions can be quite accurate and paves the way to designing robust coherence protection protocols by refocusing one interaction variations using other correlated interactions over a broad range of temperatures~\cite{wang2022coherence}.

\begin{figure}[hbtp]
\centering
\includegraphics[width=\linewidth]{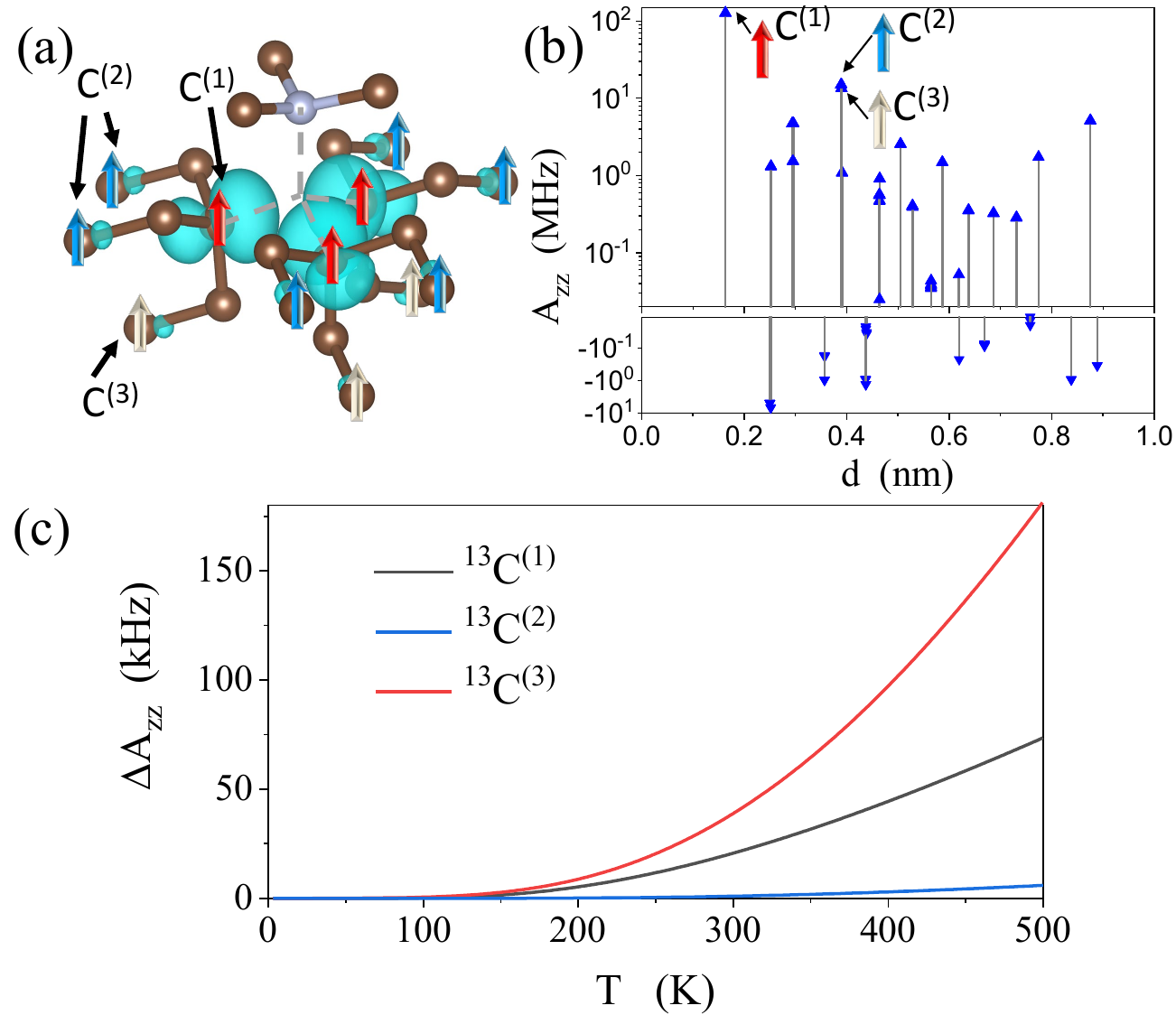}
\caption{Hyperfine interaction $A_{zz}$ with $^{13}C$ nuclear spin from simulation. (a) Illustration of the three groups of identical nuclear spins with the strongest hyperfine interaction. (b) $A_{zz}$ for different $^{13}$C nuclear spin around the NV center plotted against the distance $d$ between the $^{13}$C site and the NV center. (c) Temperature shifts of $A_{zz}$ for $^{13}$C atoms at different sites.}
\label{fig:C13}
\end{figure}

Beyond accurately reproducing experimental results, we can predict the hyperfine temperature shift of $^{13}$C nuclear spins   around the NV center, which have been actively investigated as qubits~\cite{maurer2012room,dreau2013single,barson2019temperature,shim2013room,abobeih2019atomic} (Fig.~\ref{fig:C13}a). 
%As $^{13}$C has a nuclear spin $I = 1/2$, the quadrupole interaction does not exist, and the energy splitting between $|\pm \frac{1}{2}\rangle_{m_I}$ is induced by the hyperfine interaction $A$ with electron spin. 
%As the hyperfine interaction between the NV electronic spin and $^{13}$C nuclear spin is $\vec{S}\cdot\boldsymbol{A}\cdot\vec{I}_C$, where $\vec{S}$ polarizes to $z$-direction, the hyperfine transition frequency of the $^{13}$C nuclear spin is $\nu_A = |e_z\cdot\boldsymbol{A}|/\hbar$ (without extermal magnetic field). 
%todo: we might want to explain better what we mean here!! 
% Thanks. I revised the description, which relate the 13C hyperfine transition frequency to the computed matrix A. 
$A_{zz}$ for $^{13}C$ atoms at different lattice sites around the NV centers is shown in Fig.~\ref{fig:C13}b as a function of their distance to the NV center, where we identify three groups of equivalent $^{13}$C lattice sites with the strongest interaction. The $^{13}$C atom at the first-shell lattice sites (C$^{(1)}$) of the NV center shows a strong hyperfine splitting of 129 MHz, well consistent with the experimental value of 127 MHz~\cite{barson2019temperature}. 
%We notice that the interaction strength is not a monotonic function of the distance to the NV center and is strongly related to the electron spin density distribution. 

The temperature dependence $\Delta A_{zz}(T)$ of the three groups of $^{13}$C lattice sites is shown in Fig.~\ref{fig:C13}c. $A_{zz}$ is positively related to $T$ for all the three groups of lattice sites. The most sensitive temperature dependence appears for C$^{(3)}$ lattice sites, where the temperature shift approaches 47.2 kHz at 300 K. At room temperature, the slope $dA_{zz}/dT$ for C$^{(1)}$, C$^{(2)}$, and C$^{(3)}$ are 201, 14, 440 Hz/K, respectively. 
Previously, experimental work on the temperature dependence of $^{13}$C hyperfine interaction did not provide definite results due to the high noise, while their computational study did not consider the second-order dynamical phonon effect~\cite{barson2019temperature}. Our results provide a theoretical prediction for $A_{zz}(T)$ of $^{13}$C atoms in NV centers, which could be measured by the nuclear spin transition~\cite{wang2022coherence}.

In the discussion above, we focus on the temperature shift under thermal equilibrium, but the dynamical phonon effect on spin transition frequency can be extended to kinetic processes such as phonon transport. In general, the temperature shift at a single NV center measures a linear combination $\hat{O}(r)= \sum_i\frac{\hbar}{2M_i\omega_i}\frac{\partial^2 \nu}{\partial q_i^2}\hat{n}_i(\vec{r})$ of the particle number operators $\hat{n}_i$ of ambient phonon,
where electron or nuclear spin of an NV ensemble can probe phonon distribution in the sample. While most materials characterization methods probe the atomic equilibrium position, the second-order temperature shift of spin defects provides a potential way to probe the dynamical process of local atomic vibrations.

%As the computational method does not depend on specific analysis of the NV center, 
Our work provides a general scheme to study the temperature dependence of transition frequencies of color centers. An integrated python program interfaces with VASP and phonopy will be made publicly accessible, which can carry out the numerical workflow automatically. %for input atomic configuration of color center specified by users.
Besides the NV center, various other color centers such as  SiV, SnV, PbV, GeV, MgV in diamond are also widely studied due to their potential in quantum sensing and communication~\cite{thiering2018ab,degen2017quantum,luhmann2019coulomb,pershin2021highly}. Besides diamond, point defects in silicon~\cite{zwanenburg2013silicon,hensen2020silicon,petit2018spin}, silicon-carbide~\cite{castelletto2020silicon,babin2022fabrication,anderson2022five}, Y$_2$SiO$_5$~\cite{zhong2016high}, and YVO$_4$~\cite{le2022clock,ruskuc2022nuclear} also attract broad interests. The temperature dependence data of transition frequency remains lacking for many of these systems. Applying our method to these spin defects can establish a database for temperature-dependent transition frequency of electronic excitation, electron and nuclear spin transitions, which provides critical information for searching and designing high-performance quantum devices such as highly-sensitive sensors and long-lived memories. 
%For example, a high-accuracy nanothermometry can be selected from the color center with ZFS highly sensitive to temperature. One can also explore other engineering degrees of freedom like applying load to the color center and searching for the optimal sensitivity in the strain space by calculations. 
Thus, our result paves the way to perform a systematic study on the energy levels of point defects targeted for different quantum applications.

%\acknowledgements
We thank Haowei Xu, Boning Li, and Yixuan Song for insightful discussions. This work was supported by HRI-US, NSF DMR-1923976, NSF DMR-1923929 and NSF CMMI-1922206. A.R.B. acknowledges support from a National Science Foundation Graduate Research Fellowship under Grant No. DGE-174530. The calculations in this work were performed in part on the Texas Advanced Computing Center (TACC) and MIT engaging cluster.

\bibliography{bibliography}% Produces the bibliography via BibTeX.

\clearpage
\pagebreak
\setcounter{section}{0}
\setcounter{equation}{0}
\setcounter{figure}{0}
\setcounter{table}{0}
\setcounter{page}{1}
\makeatletter
\renewcommand{\theequation}{S\arabic{equation}}
\renewcommand{\thesection}{S\arabic{section}}
\renewcommand{\thefigure}{S\arabic{figure}}
%\renewcommand{\bibnumfmt}[1]{[S#1]}
%\renewcommand{\citenumfont}[1]{S#1}

%%%%%%%%%% Prefix a "S" to all equations, figures, tables and reset the counter %%%%%%%%%%

\title{Supplemental Materials: First-Principles Calculation of the Temperature-dependent Transition Energies in Spin Defects} 

\author{Hao Tang}
\thanks{These authors contributed equally.}
\affiliation{Department of Materials Science and Engineering, Massachusetts Institute of Technology, MA 02139, USA}

\author{Ariel Rebekah Barr}
\thanks{These authors contributed equally.}
\affiliation{Department of Materials Science and Engineering, Massachusetts Institute of Technology, MA 02139, USA}

\author{Guoqing Wang}
\thanks{These authors contributed equally.}
\affiliation{
   Research Laboratory of Electronics, Massachusetts Institute of Technology, Cambridge, MA 02139, USA}
\affiliation{
   Department of Nuclear Science and Engineering, Massachusetts Institute of Technology, Cambridge, MA 02139, USA}

\author{Paola Cappellaro}\email{pcappell@mit.edu}
\affiliation{
   Research Laboratory of Electronics, Massachusetts Institute of Technology, Cambridge, MA 02139, USA}
\affiliation{
   Department of Nuclear Science and Engineering, Massachusetts Institute of Technology, Cambridge, MA 02139, USA}
\affiliation{Department of Physics, Massachusetts Institute of Technology, Cambridge, MA 02139, USA}

\author{Ju Li}\email{liju@mit.edu}
\affiliation{
   Department of Nuclear Science and Engineering, Massachusetts Institute of Technology, Cambridge, MA 02139, USA}
\affiliation{Department of Materials Science and Engineering, Massachusetts Institute of Technology, MA 02139, USA}

\maketitle

\begin{widetext}

\section{Details of first-principles calculation}
The electronic structure calculation employs the projector-augmented-wave (PAW) method implemented by the Vienna ab-initio simulation package (VASP) with a cut-off energy of 520 eV~\cite{kresse1996efficient,kresse1999ultrasoft}. 
%In the atomic relaxation and DFPT calculation, spin-unrestricted calculations are implement using the generalized gradient approximation (GGA) with the Perdew-Burke-Ernzerhof (PBE) functional for electron exchange-correlation~\cite{perdew1996generalized}. The high-accuracy hybrid functional HSE06~\cite{heyd2003hybrid} is used in the calculation of $D$, $Q$, and $A$ matrix. 
Spin-unrestricted calculations are implement using the generalized gradient approximation (GGA) with the Perdew-Burke-Ernzerhof (PBE) functional for electron exchange-correlation~\cite{perdew1996generalized} in the calculation of atomic relaxation, density functional perturbation theory (DFPT), $D$, $Q$, and $A$ matrix. 
The $k$-point mesh is sampled by the Monkhorst-Pack method~\cite{monkhorst1976special} with a seperation of 0.2 rad/\AA$^{-1}$ ($3\times 3\times 3$ $k$-point mesh in the supercell). The energy of electronic iterations converge to $10^{-7}$ eV and force on atoms converge to $0.01$ eV/\AA .

The phonon calculation is implemented in the $4\times 4\times 4$ rhombohedral supercell with only $\Gamma$ $k$-point. 381 phonon modes are obtained at $\Gamma$ $k$-point, including 3 trivial modes corresponding to overall translation. The trivial modes have no contribution to Eq.~(5) in the main text, as the overall translation of the system does not change any transition energy. All the 378 non-trivial phonon modes are then used to calculate the second-order derivative according to Eq.~(5) in the main text. In Eq.~(5), the step of displacement $\delta q_i$ is set as 0.1 \AA \ in the NV center calculation. We tested that using $\delta q_i = 0.05$ \AA \ gives almost the same temperature dependence, confirming the convergence of our results to $\delta q_i$. We mention that numerically, $\delta q_i = $ 0.1 \AA \ is not a large value as it appears to be, as this magnitude of displacement is distributed to all the 127 atoms in the supercell, and the displacement of each atom is in a proper range.  

\section{Calculation of $D$, $Q$, and $A$}
The first term in Eq.~(6) in the main text originates from the electron spin-spin magnetic dipolar interaction from the Kohn-Sham orbitals~\cite{ivady2014pressure}:
\begin{equation}
    D_{ij}=\frac{\mu_{0} g_{\rm e}^{2}\mu_{\rm B}^{2}}{4\pi} \sum_{a<b} \chi_{ab}\left\langle\Psi_{ab}\left|\frac{r^{2} \delta_{ij}-3 r_{i} r_{j}}{r^{5}}\right| \Psi_{ab}\right\rangle
\end{equation}
where $\mu_0$, $\mu_{\rm B}$, $g_{\rm e}$ are the magnetic constant, Bohr magneton, and Landé factor of electron, respectively. The summation goes through all electron pairs $(a,b)$. $\Psi_{ab}$ are particle determinant wave functions from the Kohn-Sham ground state and $\chi_{ab}$ equals +1/-1 when the states $a$ and $b$ have the same/different spin. 

The hyperfine constant (the 2$^{\rm nd}$ term in Eq.~(6) in the main text) is a summation of the Fermi contact contribution and dipolar contribution:
\begin{equation}
\begin{aligned}
    A_{ij} &= \frac{\mu_{0} g_{\rm e} g_{I}\mu_{\rm B}\mu_I}{\left\langle S_{z}\right\rangle} \times [ \frac{2}{3} \delta_{i j}  \rho_{s}\left(\mathbf{R}_{I}\right) + \\
    & \frac{1}{4 \pi} \int \frac{\rho_{s}\left(\mathbf{r}+\mathbf{R}_{I}\right)}{r^{3}} \frac{3 r_{i} r_{j}-\delta_{i j} r^{2}}{r^{2}} d \mathbf{r}]
\end{aligned}
\end{equation}
where $g_I$, $\mu_I$, and $R_I$ are the g-factor, nuclear magneton, and coordinate of nuclear spin, $\rho_s$ is the electron spin density from the DFT ground state. In the NV center calculation, we use the g-factor of $g_I = 3.077\times \frac{h}{\mu_B}$ MHz/T for $^{14}$N and $g_I = 10.7084\times \frac{h}{\mu_B}$ MHz/T for $^{13}$C.

The nuclear quadrupole interaction is calculated through the electric field gradient (EFG) $V_{ij}=\partial_iE_j$ from the electron density $\rho (r)$:
\begin{equation}
\begin{aligned}
    V_{ij} = \frac{e}{4\pi \epsilon_0}[-\int \rho (r+R_I)\frac{3 r_{i} r_{j}-\delta_{i j} r^{2}}{r^{5}}dr +\\
    \sum_{I'}Z_{I'}\frac{3 R_{II',i} R_{II',j}-\delta_{i j} R_{II'}^{2}}{R_{II'}^{5}}]
\end{aligned}
\end{equation}
where $e$, $\epsilon_0$, $Z_I'$, $R_I$, and $R_{II'}$ are the electron charge, vacuum dielectric constant, nuclear charge, position of nucleus, and relative coordinate between nucleus $I'$ and $I$. The $Q$ matrix is then calculated by the three principal axis system (PAS) eigenvalues $V_{1,2,3}$ of the EFG:  
\begin{equation}
    Q_{ij} = \frac{eQ_I}{4I(2I-1)}
  \left[ {\begin{array}{ccc}
    V_1-V_2-V_3 & 0 & 0\\
    0 & -V_1+V_2-V_3 & 0\\
    0 & 0 &  2V_3\\
  \end{array} } \right]
\end{equation}
where $Q_I$ is the nuclear quadrupole moment. In the NV center calculation, we use the $^{14}$N nuclear quadrupole moment $Q_I = 20.44$ $e\cdot$mb.

In the principal axis coordinates, Eq.~(6) in the main text is converted to:
\begin{align}
    \hat{H}_D = D[S_z^2-\frac{1}{3}S(S+1)+\frac{\epsilon}{3}(S_+^2+S_-^2)]\\
    \hat{H}_Q = Q[I_z^2-\frac{1}{3}I(I+1)+\frac{\eta}{3}(I_+^2+I_-^2)]\\
    \hat{H}_A = A_{zz}S_zI_z+A_{xx}S_xI_x+A_{yy}S_yI_y
\end{align}
where $D= \frac{3}{2}D_{zz}$ and $Q = \frac{3}{2}Q_{zz}$ are the splitting energy, and $\epsilon = (D_{xx}-D_{yy})/D_{zz}$ and $\eta = (Q_{xx}-Q_{yy})/Q_{zz}$ are the asymmetric coefficients. Applying to an uniaxial system like NV-center under zero strain, we select the $C_3$ axis as $z$-direction and the effective Hamiltonian reduces to  (considering $H_D \gg H_A, H_Q$) Eq.~(7) in the main text.

\section{Calculation of the temperature-dependent ZPL}
Although the ZPL is not spin-transition, its temperature dependence can be calculated by the same theoretical framework. The only difference is that the energy levels of the excited state and ground state have a large separation of 1.945 eV, so their difference can no longer be evaluated perturbatively. Compared with the spin transition where the transition energy is far smaller than the phonon energy, here the electronic transition energy is far larger than the phonon energy. 

%\begin{figure}[hbtp]
%\centering
%\includegraphics[width=0.4\textwidth]{sm_fig.png}
%\caption{Illustration of the transition energy of the zero phonon line considering the dynamical phonon effects.}
%\label{fig:zpl}
%\end{figure}

The calculation is implemented through evaluating Eq.~(4) in the main text. At first, we use the $\Delta$ SCF method~\cite{PhysRevB.78.075441} to calculate the relaxed atomic structure and electronic structures of the excited states, where one electron is placed in a high-lying Kohn-Sham orbital. In the NV center calculation, this is realized by fixing the electron occupancy according to the $^3E$ excited state's $a_1e^3$ molecular orbital configuration as in ref.~\cite{gali2009theory}. (The same electron occupancy is set for all $k$ points in the supercell calculation, yielding the $^3E$ state with one excited electron in each NV center, or namely, one excited electron in each supercell). The phonon spectral of the excited states is then calculated, deriving a series of excited state phonon frequency $\omega^e_i$. Combining with the ground state phonon frequency $\omega^g_i$, we derive $\Delta \omega_i$ in Eq.~(4) in the main text. $\omega_i$ in the equation is set as the ground state phonon frequency, as phonon exhibits the Boson distribution according to the ground state phonon spectral in the ZPL measurement. The first-order contribution from $\nu_0(T)$ is directly calculated through the excitation energy calculation under lattice expansion by the $\Delta$ SCF method.

\section{Convergence test of the supercell size}
\begin{figure}[hbtp]
\centering
\includegraphics[width=0.6\linewidth]{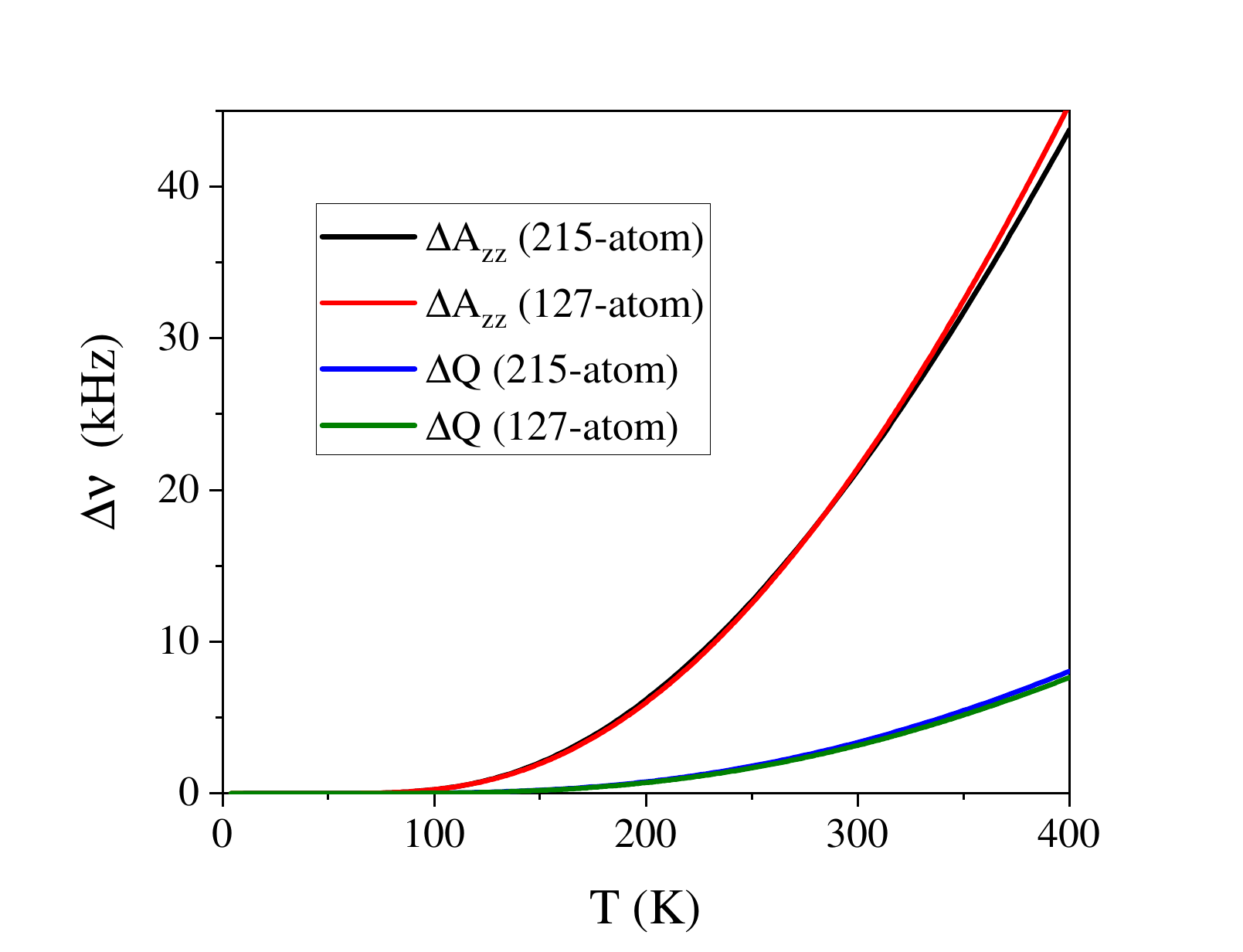}
\caption{Convergence test of the temperature dependence of $A_{zz}$ and $Q$ of the $^{14}$N nuclear spin in NV centers. The black/blue and red/green curves represent the temperature drifts of $A_{zz}$/$Q$ of the $^{14}$N atom calculated by the 215-atom rhombohedral supercell and 127-atom cubic supercell, respectively}
\label{fig:converge}
\end{figure}
Here we conducted convergence test to show that the 127-atom $4\times 4\times 4$ rhombohedral supercell we used for NV centers can derive temperature dependence reasonably convergent against supercell size. The temperature dependence of $A_{zz}$ and $Q$ is calculated by both the 127-atom supercell and a 216-atom cubic supercell, as shown in Fig.~\ref{fig:converge}. The temperature dependence calculated by different supercell is well consistent, supporting that the 127-atom supercell can derive temperature dependence with reasonable convergence. As the computational cost of the temperature dependence calculation grows dramatically with the supercell size, we recommend the 127-atom supercell based on our test results to people who intend to use our method. 

\end{widetext}

\end{document}